\documentclass{article}[10pt]
\usepackage{amsmath}
\usepackage{amssymb}
\usepackage{amsfonts}
\usepackage{physics}
\usepackage{siunitx}
\usepackage{mhchem}
\usepackage{hyperref}
\usepackage{graphicx}
\usepackage{geometry}[margin=0.2cm]
\usepackage{tikz}
\usepackage{setspace}
\usetikzlibrary{shapes.geometric}
\usetikzlibrary{arrows.meta}
\usetikzlibrary{calc}
\usetikzlibrary{patterns}
\usetikzlibrary{fit}

\newcommand{\doubleprime}{{\prime\prime}}
\newcommand{\keff}{k_{\text{eff}}}
\providecommand{\keywords}[1]{\textbf{\textit{Keywords---}} #1}

\DeclareSIUnit\pcm{pcm}
\title{An iterative scheme to include turbulent diffusion in advective-dominated transport of delayed neutron precursors}
\author{Mathis Caprais\thanks{Université Paris-Saclay, CEA, Service d'\'{E}tudes des Réacteurs et de Mathématiques Appliquées, Gif-sur-Yvette, 91191, France}, André Bergeron\thanks{Université Paris-Saclay, CEA, Service de Thermo-hydraulique et de Mécanique des Fluides, Gif-sur-Yvette, 91191, France}}
\date{\today}
\begin{document}
\onehalfspacing
\maketitle
\begin{abstract}
    In this study, the Method of Characteristics (MOC) for Delayed Neutron Precursors (DNPs) is used to solve the precursors balance equation with turbulent diffusion. The diffusivity of DNPs, significantly higher than molecular diffusivity, emerges in turbulent flows from the time-averaging of the DNPs mass balance equation. To integrate this effect within the MOC framework, the advection-reaction component of the DNPs balance equation is solved using the MOC, while the diffusive source is computed from the concentration of the previous iteration. The method is validated on a 2D recirculating pipe reactor with high Reynolds number flow, comparing the MOC with diffusion to a standard finite volume (FV) discretization of the fission products balance equation. Additionally, the impact of the diffusivity term on DNP distributions and reactor reactivity is quantified as a function of the turbulent Schmidt number.
\end{abstract}
\section{Introduction}
A numerical method based on the Method of Characteristics (MOC) is developed to account for the diffusivity of Delayed Neutron Precursors (DNPs) in liquid nuclear fuels. Although it is commonly accepted that molecular diffusivity is negligible compared to advection in liquid fuel reactors \cite{cheng2014development, AUFIERO201478}, the turbulent diffusivity of DNPs arising in Reynolds Averaged Navier-Stokes (RANS) simulations is orders of magnitude larger than molecular diffusivity. The method presented in this work includes the diffusivity term in the DNPs balance equation while maintaining the MOC framework.

To achieve this goal, the advection-reaction part of the DNPs equation is solved using pathline coordinates, allowing for a fast numerical calculation of the concentrations for a given velocity field \cite{caprais2024new}. The diffusivity term is moved to the right-hand side of the balance equation and forms a new source term calculated with the previous concentrations. This scheme forms a fixed-point problem that can be solved with various iterative techniques, such as Picard iterations or the Jacobian-Free Newton-Krylov (JFNK) method.

This method is tested on a two-dimensional circulating liquid fuel reactor at a high Reynolds number. The reactor is modeled as a two-region system: the core with non-zero fission cross-sections and the recirculation loop simulating a possible heat exchanger with a zero fission cross-section. Although the fission cross-section is never zero, the neutron flux is negligible is this zone of the reactor. Thermal feedback is not considered in this analysis. However, since the numerical method would remain unchanged and be applied at each thermohydraulic coupling step, the presented test case is sufficient for validation.

The MOC for DNPs is compared to a finite volume implementation of the advection-diffusion-reaction equation for DNPs. The results are compared in terms of computational time and accuracy.

The diffusivity of DNPs appears as a second-order effect in the DNPs balance equation and is mostly visible for long-lived precursors. The inclusion of the diffusion term is quantified in terms of the reactor's reactivity, neutron flux, and DNPs distributions.

The paper is organized as follows: in Sec. \ref{sec:MMOC}, the MOC for DNPs is presented alongside the advection-diffusion-reaction equation and its dimensional analysis. Pathline coordinates are reviewed in Sec. \ref{subsec:pathlines_coordinates}, and the iterative method accounting for the diffusivity term is presented in Sec. \ref{subsec:splitting_diffusion_advection}. The discretized equations are presented in Sec. \ref{subsec:discretized_equations}. In Sec. \ref{subsec:neutron_balance}, the neutron balance equation is recalled, discretized, and the coupling between the MOC for DNPs and the neutron transport equation is presented. The test case is described in Sec. \ref{sec:2D_pipe}, and the results are discussed in Sec. \ref{sec:results}. Finally, the conclusions are drawn in Sec. \ref{sec:conclusion}.
\section{The Method of Characteristics for DNPs}\label{sec:MMOC}
Before solving the DNPs equation with the MOC, the general advection-diffusion-reaction equation is recalled in Sec. \ref{subsec:advection_diffusion}. This balance equation is averaged in Sec. \ref{subsubsec:RANS_averaging} to obtain a steady-state version of the DNPs balance with a turbulent diffusivity coefficient. This balance equation is then scaled in Sec. \ref{subsubsec:nondimensionalization} to highlight its dominant terms. Pathlines coordinates are reminded in Sec. \ref{subsec:pathlines_coordinates}. Diffusivity is included in the MOC in Sec. \ref{subsec:splitting_diffusion_advection}. The discretized equations are presented in Sec. \ref{subsec:discretized_equations}, the pathline tracking in Sec. \ref{subsec:pathlines_tracing}. Finally, the calculation strategy is presented in Sec. \ref{subsec:calculation_strategy}.
\subsection{Advection-Diffusion Equation}\label{subsec:advection_diffusion}
\subsubsection{Precursors balance equation}\label{subsec:precursors_balance}
In a liquid fuel, the concentration of fission products such as DNPs follows a mass transport equation which is a balance between advection, molecular and turbulent diffusion, decay and production by a source. For the $j$-family of DNPs with a concentration $C_j$ (\SI{}{\per\cubic\meter}),
\begin{equation}
    \pdv{C_j}{t} + \vb{u}\cdot\grad{C_j} - \div{D\grad{C_j}}+ \lambda_j C_j = S_j,
    \label{eq:precursors_advection_diffusion}
\end{equation}
where $\vb{u}$ (\SI{}{\meter\per\second}) is the velocity field, $D$ (\SI{}{\square\meter\per\second}) is the molecular diffusivity, $\lambda_j$ (\SI{}{\per\second}) is the decay constant of the $j$-family of DNPs and $S_j$ (\SI{}{\per\cubic\meter\per\second}) is the source term. The source term $S_j$ is a function of the fission rate and the fission yield of the $j$-family of DNPs.
\subsubsection{Reynold Averaging of the DNPs balance equation}\label{subsubsec:RANS_averaging}
The effective diffusivity coefficient follows from the RANS formalism. The velocity field and the concentration are decomposed into a sum of a mean and a fluctuating temporal part, and the original balance equation Eq. \eqref{eq:precursors_advection_diffusion} is averaged over a period $T \to \infty$ to obtain a steady-state balance equation for the DNPs concentration \cite{tong_fluid_mechanics, Pope_2000}:
\begin{equation}
    \vb{u}\cdot\grad{C_j} - \div{D_t\grad{C_j}} + \lambda_j C_j = S_j
    \label{eq:precursors_advection_diffusion_RANS_turbulent}
\end{equation}
where the turbulent diffusivity $D_t$ is defined as:
\begin{equation}
    D_t = \frac{\nu_t}{\text{Sc}_t}.
    \label{eq:turbulent_diffusivity}
\end{equation}
The turbulent kinematic viscosity is denoted as $\nu_t$, and the turbulent Schmidt number is represented as $\text{Sc}_t$. In some works, both turbulent and molecular diffusion are considered; but given that molecular diffusion ($D$) is often significantly smaller than turbulent diffusion ($D_t$), molecular diffusion is neglected in the balance equation \cite{cheng2014development, AUFIERO201478}.
\subsubsection{Dimensional Analysis}\label{subsubsec:nondimensionalization}
To better understand the dominant terms in the DNPs balance equation (Eq. \eqref{eq:precursors_advection_diffusion_RANS_turbulent}), we perform a dimensional analysis. This involves scaling the variables to highlight the relative importance of different physical processes. Scaling involves introducing dimensionless variables:
\begin{itemize}
    \item \( \vb{u}^\prime = \vb{u} / u_0 \): Dimensionless velocity, where \( u_0 \) (\SI{}{\meter\per\second}) is a characteristic velocity.
    \item \( \vb{r}^\prime = \vb{r} / L \): Dimensionless position, with \( L \) (\SI{}{\meter}) being a characteristic length scale.
    \item \( C_j^\prime = C_j / C_0 \): Dimensionless concentration, where \( C_0 \) (\SI{}{\per\cubic\meter}) is a reference concentration.
    \item \( S_j^\prime = S_j / S_0 \): Dimensionless source term, with \( S_0 \) (\SI{}{\per\cubic\meter\per\second}) as a reference source term.
    \item \( D^\prime = D_t / D_0 \): Dimensionless turbulent diffusivity, with \( D_0 \) (\SI{}{\square\meter\per\second}) being a characteristic diffusivity.
\end{itemize}
Since the DNPs source term \( S_j \) is related to the neutron flux, which itself depends linearly on the precursor concentrations, we have \( S_0 \propto \lambda_j C_0 \). Substituting these dimensionless variables into the balance equation, we get:
\[
    \vb{u}^\prime\cdot\grad{C_j^\prime} - \frac{1}{\mathcal{A}}\div{D^\prime\grad{C_j^\prime}}+  \frac{1}{\mathcal{B}_j}C_j^\prime = \frac{S_j^\prime}{\mathcal{B}_j},
\]
with scaling parameters defined as:
\begin{equation}
    \mathcal{A} = \frac{L u_0}{D_0}, \quad \mathcal{B}_j = \frac{u_0}{L \lambda_j} \qq{and} \mathcal{C}_j = \mathcal{B}_j / \mathcal{A}. 
    \label{eq:adim_parameters}
\end{equation}
The parameter \( \mathcal{A} \) is the ratio of advection to diffusion, akin to the inverse of the Péclet number. The parameter \( \mathcal{B}_j \) represents the ratio of advection timescale to the precursor decay timescale. A higher \( \mathcal{A} \) indicates that advection dominates over diffusion, while a higher \( \mathcal{B}_j \) suggests that advection occurs much faster than the decay of DNPs. These dimensionless numbers can be composed to form the number \( \mathcal{C}_j \), which is the ratio of the diffusion timescale to the decay timescale.

From these dimensionless numbers, we can infer that long-lived precursors, i.e. those with the lowest decay constant are more affected by both turbulent diffusion and advection.
\subsection{Pathlines coordinates}\label{subsec:pathlines_coordinates}
Pathlines represent the trajectories that fluid particles follow within a flow field and serve primarily as a visualization tool. In the context of a steady-state velocity field, particularly within the Method of Characteristics (MOC) framework, pathlines correspond to the integral curves of the velocity field. They are also the characteristics of the advection balance equations. The pathlines equation of motion is,
\begin{equation}
    \dv{\vb{r}}{\tau} = \vb{u}\qty(\vb{r}) \qq{with} \vb{r}(0) = \vb{r}_0,
    \label{eq:pathlines}
\end{equation}
where $\tau$ is called the time-of-flight and is the time spent by a particle along the trajectory and $\vb{r}_0$ is the initial position of the fluid particle. These particles are passive as they do not interact back with the flow. As the fluid is assumed to be divergence-free,
\begin{equation}
    \div{\vb{u}} = 0,
    \label{eq:incompressible}
\end{equation}
a potential vector of the velocity field $\vb{u}$ can be defined and represented as \(\vb{u} = \curl{\vb*{\Pi}} \qq{and} \vb*{\Pi} = \grad{\varphi} + \chi\grad{\psi}\), where $\varphi$, $\chi$ and $\psi$ are scalar fields and usually called Monge or Clebsch potentials. This representation of the potential vector yields a new representation of the velocity field. This allows writing the velocity field as \(\vb{u} = \grad{\chi} \crossproduct \grad{\psi}\), which is its representation in terms of potentials. The triplet $\qty(\tau, \chi, \psi)$ is a new set of coordinates that can be used to describe the flow. Along a pathline, both $\chi$ and $\psi$ are constant as their gradients are orthogonal to the velocity field $\qty(\vb{u}\cdot\grad{\chi} = \vb{u}\cdot\grad{\psi}=0)$ due to the representation of \(\vb{u}\) being a cross-product of potential gradients. The potentials introduced in the velocity field are known as the bistream function and induce a change of coordinate that conserves the infinitesimal volume element $\dd{V} = \dd{x}\dd{y}\dd{z} = \dd{\tau}\dd{\chi}\dd{\psi}$, which is a consequence of the incompressibility of the flow, Eq. \eqref{eq:incompressible}. In the scope of this work, the bistream function identifies as the well-known streamfunction \cite{ferziger2019computational}.
\subsection{MOC formalism and advection-diffusion splitting}\label{subsec:splitting_diffusion_advection}
\subsubsection{MOC equations for DNPs}\label{subsubsec:MOC_formalism}
To account for DNPs diffusion in the MOC method, the balance equation Eq. \eqref{eq:precursors_advection_diffusion_RANS_turbulent} is splitted. The reason for splitting the equations is that, once the pathlines kinematic equation (Eq. \eqref{eq:pathlines}) is integrated, the advection-reaction part of the balance equation can be easily inverted. The linear problem described by Eq. \eqref{eq:precursors_advection_diffusion_RANS_turbulent} forms a system of equations represented as \( A_j \vb{x}_j = \vb{S}_j \), where \( A_j = B_j - C \). Here, \( A_j \) is a linear operator split into an advection-reaction operator \( B_j \) and a diffusion operator \( C \), \( \vb{x}_j \) is the DNPs concentration vector, and \( \vb{S}_j \) is the source term. The MOC effectively inverts the advection-reaction operator \( B_j = \vb{u} \cdot \nabla + \lambda_j \) with a known source term. By treating the problem as a fixed-point, the solution is found iteratively as 
\[
\vb{x}_j^{(m)} = B_j^{-1} \left( C \vb{x}_j^{(m-1)} + \vb{S}_j \right).
\]
The new source term is defined as,
\begin{equation}
    S_j^{(m-1)} \equiv S_j + S_{\mathrm{diff}}^{(m-1)} = S_j + \div{D_t\grad{C_j^{(m-1)}}},
    \label{eq:source_term_MMOC}
\end{equation}
and encompasses the turbulent diffusion term evaluated with the concentration of the \((m-1)\) fixed point iteration. The advection term of Eq. \eqref{eq:precursors_advection_diffusion_RANS_turbulent} is rewritten using pathline coordinates,
\begin{equation}
    \dv{}{\tau}C_j^{(m)}\qty(\vb{r}\qty(\tau))  + \lambda_j C_j^{(m)}\qty(\vb{r}\qty(\tau))  = S_j^{(m-1)}\qty(\vb{r}\qty(\tau)),
    \label{eq:precursors_advection_pathlines_ode}
\end{equation}
effectively transforming the partial differential equation into an ordinary differential equation of which the analytical solution is:
\begin{equation}
    C_j^{(m)} \qty(\vb{r}\qty(\tau))
    = C_j^{(m)}\qty(\vb{r}\qty(\tau_{0}))e^{-\lambda_j\qty(\tau -\tau_{0})}
    + \int_{\tau_{0}}^{\tau}\dd{\tau^\prime} S_j^{(m-1)} \qty(\vb{r}\qty(\tau^\prime))e^{-\lambda_j\qty(\tau - \tau^\prime)} \qq{and} \tau_{0} < \tau.
    \label{eq:balance_precs_sol_incomp}
\end{equation}
Eq. \eqref{eq:balance_precs_sol_incomp} convoluates the modified DNPs source term of the previous iteration with an exponential decay along the pathline. In this paper, this linear fixed-point problem is solved using the JFNK method \cite{KNOLL2004357}.
\subsection{Discretized equations}\label{subsec:discretized_equations}
To effectively use the solution along a pathline, Eq. \eqref{eq:balance_precs_sol_incomp}, the source is assumed to be constant within the mesh cell $k$ for a given pathline, $S_j^{(m-1)}\qty(\vb{r}) \equiv S_{jk}^{(m-1)}, \, \vb{r}\in V_k$. Let $\tau_k^\prime$ and $\tau_k^\doubleprime$ be the time-of-flight of the pathline within the cell $k$ between the beginning and the end of that cell. The difference of concentration at the end of the cell and at the beginning of the cell is given by,
\begin{equation}
    C\qty(\vb{r}\qty(\tau_k^\doubleprime)) - C\qty(\vb{r}\qty(\tau_k^\prime)) = \qty(\frac{S_k}{\lambda} - C\qty(\vb{r}\qty(\tau_k^\prime)))\qty(1 - \exp(-\lambda T_{k})) \qq{with} T_{k} = \tau_k^\doubleprime - \tau_k^\prime,
    \label{eq:discretized_transmission}
\end{equation}
where the DNPs index and the iteration index are omitted for clarity. The pathline-averaged concentration within the cell $k$ is,
\begin{equation}
    \expval{C}_k = \frac{1}{T_k}\int_{\tau_k^\prime}^{\tau_k^\doubleprime}\dd{\tau} C\qty(\vb{r}\qty(\tau)) = \frac{S_k}{\lambda} - \frac{C\qty(\vb{r}\qty(\tau_k^\doubleprime)) - C\qty(\vb{r}\qty(\tau_k^\prime))}{\lambda T_k}.
    \label{eq:pathline_average_concentration}
\end{equation}
The transmission equation Eq. \eqref{eq:discretized_transmission} is used to compute the pathline-averaged concentration within the cell $k$. The volume average concentration of DNPs within the cell $k$ is computed as a weighted sum of the pathline-averaged concentration within the cell $k$,
\begin{equation}
    C^{(k)} = \frac{1}{V_k}\int_{V_k}\dd{V} C = \sum_i \omega_{i} \expval{C}_i, \qq{with} \omega_{i} = \frac{q_i T_i}{\sum_i q_i T_i},
    \label{eq:volume_average_concentration}
\end{equation}
where $q_k$ is the volume flow rate associated to a pathline. The volume average concentration is then mapped onto the original mesh to compute the diffusivity term. The diffusivity term is computed as,
\begin{equation}
    S_{\mathrm{diff}} = \frac{1}{V_k}\int_{V_k}\dd{V} \div{D_t\grad{C}} = \frac{1}{V_k}\int_{\partial V_k}\dd{S} \vb{F}\cdot \vu{n},
    \label{eq:diffusivity_term}
\end{equation}
where $\vb{F} = D_t\grad{C}$ is the diffusive flux and $\vu{n}$ is the outward normal vector to the cell $k$. The diffusive fluxes are computed using finite volumes.
\subsection{Pathlines tracing}\label{subsec:pathlines_tracing}
\begin{figure}[hbtp]
    \centering
    \begin{tikzpicture}[>=Stealth]
        \coordinate (g) at (0,0);
        \coordinate (g1) at (0,4);
        \coordinate (g2) at (6,4);
        \coordinate (g3) at (6,0);

        \draw (g) -- (g1) node[midway, above, sloped, xshift=-0.5cm, yshift=0.0cm] {West};
        \draw (g1) -- (g2) node[midway, above, sloped, xshift=0.5cm, yshift=0.0cm] {North};
        \draw (g2) -- (g3) node[midway, above, sloped, xshift=-0.5cm, yshift=0.0cm] {East};
        \draw (g3) -- (g) node[midway, below, sloped, xshift=-0.5cm, yshift=-0.0cm] {South};

        \draw[->] ($(g1)!0.5!(g)$)++(-1, 0) -- +(2,0) node[above right] {$u_w$};

        \draw[->] ($(g1)!0.5!(g2)$)++(0, -1) -- +(0,2) node[above right] {$u_n$};

        \draw[->] ($(g2)!0.5!(g3)$)++(-1,0) -- +(2,0) node[above right] {$u_e$};

        \draw[->] ($(g3)!0.5!(g)$)++(0, -1) -- +(0,2) node[above right] {$u_s$};

        \node[below] at (g) {$\vb{r}_\star$};
        \node[circle, fill, inner sep=1pt] at (g) {};

        \coordinate (l1) at (3, 2);
        \node[above] at (l1) {$\vb{r}_0$};
        \node[circle, fill, inner sep=1pt] at (l1) {};

        \coordinate (l2) at (5.5, 4);
        \node[above right] at (l2) {$\vb{r}_e$};
        \node[circle, fill, inner sep=1pt] at (l2) {};
        \draw (l1) to[out=0,in=-90] (l2);

        \coordinate (l3) at (0.3, 0);
        \draw[dashed] (l1) to[out=180,in=90] (l3);

    \end{tikzpicture}
    \caption{A rectangular cell with velocity components known on the faces and a forward (backward) hyperbolic (dashed) trajectory.}
    \label{fig:first_order_velocity}
\end{figure}
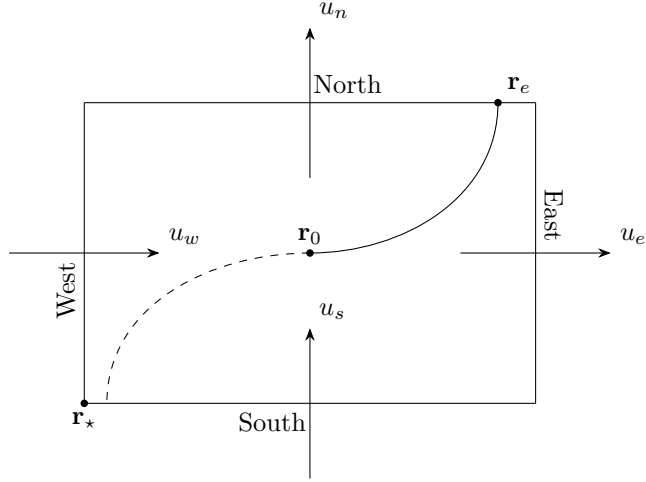

The MOC equations, Eqs. \eqref{eq:discretized_transmission}, \eqref{eq:pathline_average_concentration} and \eqref{eq:volume_average_concentration} require the knowledge of the cells crossed by pathlines, as well as their associated time-of-flight within a cell, $T_k$. The integration of the pathlines equation of motion, Eq. \eqref{eq:pathlines}, depends on the level of discretization of the velocity field at the mesh level. For the specific case of the simplified geometry of this work, the velocity field is obtained on a Cartesian mesh cell of size $(\Delta x, \Delta y)$, and velocity components are calculated on the faces of the mesh cells, as shown in Fig. \ref{fig:first_order_velocity}.
The pathline equation of motion is integrated analytically within the cell assuming a linear evolution of both the $x$ and $y$ components of the velocity field,
\begin{equation}
    u_x = u_w + \frac{u_e - u_w}{\Delta x}(x-x_\star) = u_w + \alpha_x (x-x_\star),
    \label{eq:first_order_velocity_x}
\end{equation}
and,
\begin{equation}
    u_y = u_s + \frac{u_n - u_s}{\Delta y}(y-y_\star) = u_s + \alpha_y (y-y_\star).
    \label{eq:first_order_velocity_y}
\end{equation}
In both Eqs. \eqref{eq:first_order_velocity_x} and \eqref{eq:first_order_velocity_y}, $\vb{r}_\star = \qty(x_\star, y_\star)$ serves as a reference point in the cell. Eqs. \eqref{eq:first_order_velocity_x} and \eqref{eq:first_order_velocity_y} are integrated with respect to $\tau$ to obtain the trajectory of the pathline within the cell, starting from a point $\vb{r}_0$ at time $\tau = 0$ within the cell:
\begin{equation}
    \vb{r}\qty(\tau) = \vb{r}_0
    + \begin{pmatrix}
        \frac{\qty(u_w + \alpha_x (x_0 - x_\star))\qty(\exp(\alpha_x \tau) - 1)}{\alpha_x} \\
        \frac{\qty(u_s + \alpha_y (y_0 - y_\star))\qty(\exp(\alpha_y \tau) - 1)}{\alpha_y}
    \end{pmatrix}.
    \label{eq:P1_trajectory}
\end{equation}
The trajectory equation, Eq. \eqref{eq:P1_trajectory} is then used to compute the time-of-flights required to reach a cell boundary. For $x$-boundaries,
\begin{equation}
    \tau_{w} = \frac{1}{\alpha_x}\ln(\frac{u_w}{u_w + \alpha_x (x_0 - x_\star)}) \qq{and} \tau_e = \frac{1}{\alpha_x}\ln(\frac{u_e}{u_w + \alpha_x (x_0 - x_\star)}).
    \label{eq:first_order_tof_x}
\end{equation}
The time-of-flights to reach a $y$-boundary are,
\begin{equation}
    \tau_s = \frac{1}{\alpha_y}\ln(\frac{u_s}{u_s + \alpha_y (y_0 - y_\star)}) \qq{and} \tau_n = \frac{1}{\alpha_y}\ln(\frac{u_n}{u_s + \alpha_y (y_0 - y_\star)}).
    \label{eq:first_order_tof_y}
\end{equation}
In the case where the velocity field is constant for at least one spatial axis, a Taylor expansion of the trajectory equation, Eq. \eqref{eq:P1_trajectory} with $\alpha_x \to 0$ and/or $\alpha_y \to 0$ shows that the trajectory reverts to a straight line within the direction considered. The calculation of the time-of-flights becomes straightforward. The time-of-flight to reach a $y$-boundary is,
\begin{equation}
    \quad \tau_{w} = \frac{x_\star - x_0}{u_x} \qq {and}\tau_{e} = \frac{x_\star + \Delta x - x_0}{u_x}.
    \label{eq:time_of_flight_x_oriented_P0}
\end{equation}
and to reach an $x$-boundary,
\begin{equation}
    \tau_{s} = \frac{y_\star - y_0}{u_y} \qq{and} \tau_{n} = \frac{y_\star + \Delta y - y_0}{u_y}.
    \label{eq:time_of_flight_y_oriented_P0}
\end{equation}
These results can also be obtained by a Taylor expansion of the time-of-flight equations, Eqs. \eqref{eq:first_order_tof_x} and \eqref{eq:first_order_tof_y} with $\alpha_x \to 0$ and/or $\alpha_y \to 0$.
\subsection{Calculation strategy}\label{subsec:calculation_strategy}
The calculation strategy begins by utilizing a known source of DNPs coming from the power iteration method which is used in the MOC equations, Eqs. \eqref{eq:discretized_transmission}, \eqref{eq:pathline_average_concentration} and \eqref{eq:volume_average_concentration}.

Before starting the coupled neutronic calculations, which account for DNPs advection and diffusion, pathlines are traced within the geometry using a steady-state Reynolds-Averaged Navier-Stokes (RANS) velocity field. The objective is to ensure complete coverage, meaning that every cell in the geometry is crossed by at least one pathline.

An initial estimate of the DNPs concentration is obtained by solving the balance equation (Eq. \eqref{eq:precursors_advection_diffusion_RANS_turbulent}) without considering diffusion. Once the initial concentration distributions are available, the diffusivity term (Eq. \eqref{eq:diffusivity_term}) is calculated using finite difference methods. The DNPs source term is then updated using Eq. \eqref{eq:source_term_MMOC}.

\begin{figure}[hbtp]
    \centering
    \begin{tikzpicture}[node distance=2cm and 2cm, auto, scale=0.8, transform shape]
        \tikzstyle{startstop} = [rectangle, rounded corners, minimum width=3cm, minimum height=1cm,text centered, draw=black, fill=red!30]
        \tikzstyle{process} = [rectangle, minimum width=3cm, minimum height=1cm, text centered, text width=4cm, draw=black, fill=orange!30]
        \tikzstyle{decision} = [rectangle, minimum width=5.5cm, minimum height=1cm, text centered, text width=5cm, draw=black, fill=green!30]
        \tikzstyle{arrow} = [thick,->,>=stealth]
        \node (start) [startstop] {Start};
        \node (pathlines_tracing) [process, below of=start] {Pathlines tracing};
        \node (adv_calc) [process, below of=pathlines_tracing] {Pure advection calculation};
        \node (diff_calc) [process, below of=adv_calc] {Calc. diffusive fluxes};
        \node (source_update) [process, below of=diff_calc] {Update source term};
        \node (SOR) [process, below of=source_update] {Calculate $C_j^{(m)}$};
        \node (convergence) [decision, below of=SOR] {$\norm{C_j^{(m)} - C_j^{(m-1)}} / \norm{C_j^{(m)}} < \varepsilon$};
        \node (end) [startstop, below of=convergence] {End};

        \node [draw=black, thick, dashed, inner sep=0.5cm, fit=(diff_calc) (convergence)] (iteration_box) {};

        \coordinate[left of=convergence, xshift=-1cm] (leftconverged);
        \coordinate[left of=diff_calc, xshift=-1cm] (leftdiffcalc);
        \draw [arrow] (start) -- (pathlines_tracing);
        \draw [arrow] (pathlines_tracing) -- (adv_calc);
        \draw [arrow] (adv_calc) -- (diff_calc);
        \draw [arrow] (diff_calc) -- (source_update);
        \draw [arrow] (source_update) -- (SOR);
        \draw [arrow] (SOR) -- (convergence);
        \draw (convergence) -- (leftconverged);
        \draw (leftconverged) -- node[midway, sloped, below] {no} (leftdiffcalc);
        \draw [arrow] (leftdiffcalc) -- (diff_calc);
        \draw [arrow] (convergence) -- node[anchor=east] {yes} (end);
    \end{tikzpicture}
    \caption{Flowchart of the DNPs calculation method, the dashed box corresponds to the fixed-point iteration within the JFNK method.}
    \label{fig:flowchart}
\end{figure}
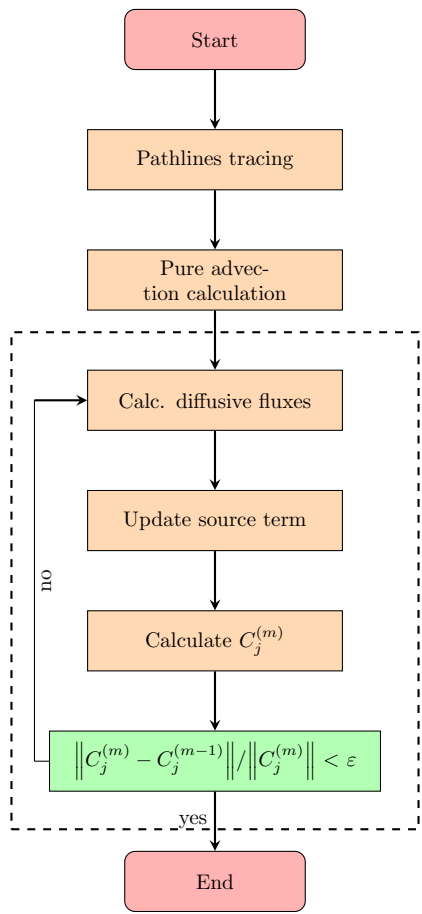

To solve the fixed-point problem, we employ the Jacobian-Free Newton-Krylov (JFNK) method. This choice is made because the conventional Successive Over Relaxation (SOR) method shows a slow convergence rate for the application studied in this paper. While SOR often results in linear convergence, the JFNK method typically provides superlinear convergence \cite{KNOLL2004357}. The entire calculation method is summarized in the flowchart presented in Fig. \ref{fig:flowchart}.

The precursor calculation scheme, Fig. \ref{fig:flowchart}, is called at each power iteration when the DNPs source is modified.
\subsection{Finite volumes implementation}\label{subsec:finite_volumes}
To verify the iterative method presented in Sec. \ref{subsec:splitting_diffusion_advection}, the balance equation Eq. \eqref{eq:precursors_advection_diffusion_RANS_turbulent} is discretized using a finite volume method. The balance equation Eq. \eqref{eq:precursors_advection_diffusion} is integrated over a cell of size $\Delta x \times \Delta y$ to obtain the finite volume equation. For convenience, the DNPs family index is dropped and the discretized advection term is:
\begin{equation}
    \frac{1}{\Delta x \Delta y}\int_{x_{i-1/2}}^{x_{i+1/2}}\dd{x}\int_{y_{j-1/2}}^{y_{j+1/2}}\dd{y} \div{\vb{u}C} = \frac{u_{i+1/2,j}C_{i+1/2,j} - u_{i-1/2,j}C_{i-1/2,j}}{\Delta x},
    \label{eq:advection_finite_volume}
\end{equation}
where $u_{i\pm 1/2,j}$ is the $x$-velocity component at the face $i\pm 1/2$ and $j$. The velocity components along $y$ are equal to zero. To relate the face values to the cell values, the upwind scheme is used, and the face values are computed as \(u_{i+1/2,j}C_{i+1/2,j} \simeq u_{i,j}C_{i,j}\) and \(u_{i-1/2,j}C_{i-1/2,j} \simeq u_{i-1,j}C_{i-1,j}\). Eq. \eqref{eq:advection_finite_volume} forms a bidiagonal system of equations. The decay term of Eq. \eqref{eq:precursors_advection_diffusion} is discretized as,
\begin{equation}
    \frac{1}{\Delta x \Delta y}\int_{x_{i-1/2}}^{x_{i+1/2}}\dd{x}\int_{y_{j-1/2}}^{y_{j+1/2}}\dd{y} \lambda C = \lambda C_{i,j}.
    \label{eq:decay_finite_volume}
\end{equation}
Eq. \eqref{eq:decay_finite_volume} is a diagonal matrix. The diffusivity term is discretized as,
\begin{equation}
    \begin{aligned}
        \frac{1}{\Delta x \Delta y}
        \int_{x_{i-1/2}}^{x_{i+1/2}} \dd{x}
        \int_{y_{j-1/2}}^{y_{j+1/2}} \dd{y}
        \div{D \grad{C}}
         & \simeq
        \frac{1}{\Delta x \Delta y}           \\
         & \quad \times \Big[
            \delta_{i-1,j}(C_{i-1,j} - C_{i,j})
        + \delta_{i+1,j}(C_{i+1,j} - C_{i,j}) \\
         & \quad \quad
            + \delta_{i,j-1}(C_{i,j-1} - C_{i,j})
            + \delta_{i,j+1}(C_{i,j+1} - C_{i,j})
            \Big],
    \end{aligned}
    \label{eq:diffusivity_finite_volume}
\end{equation}
where the turbulent index $t$ is dropped for clarity. Eq. \eqref{eq:diffusivity_finite_volume} is a pentadiagonal matrix, and $\delta_{i+1,j}$ is the harmonic average of the diffusivity coefficients, \( \delta_{i+1,j} = 2D_{i,j}D_{i+1,j} / \qty(D_{i,j} + D_{i+1,j}) \) following from the conservation of Fick's law at the interfaces between cells.
The system of equations formed by Eqs. \eqref{eq:advection_finite_volume}, \eqref{eq:decay_finite_volume} and \eqref{eq:diffusivity_finite_volume} with their associated boundary conditions is solved using Scipy sparse linear algebra solver \texttt{scipy.sparse.linalg.spsolve} \cite{virtanen2020scipy}.
\section{Coupling DNPs transport with neutron transport}\label{subsec:neutronics_simulation}
The DNPs calculation is coupled to a neutron flux calculation through the DNPs source of Eq. \eqref{eq:precursors_advection_diffusion_RANS_turbulent}. The multigroup neutron transport equation is reminded in Sec. \ref{subsec:neutron_balance}. This equation is discretized in angle in Sec. \ref{subsec:angular_discretization} and in space in Sec. \ref{subsec:space_discretization}. The iterative scheme to solve the coupled DNPs and neutron transport problem over a periodic $2$D domain is presented in Sec. \ref{subsec:iterative_scheme}.
\subsection{Multigroup neutron balance equation}\label{subsec:neutron_balance}
In the steady state, the multigroup neutron transport equation is a balance equation between streaming, absorption, scattering and production,
\begin{equation}
    \vb*{\Omega}\cdot\grad{\psi_g} + \Sigma_t^g \psi_g = \mathcal{S}_g\psi + \frac{1}{\keff}\mathcal{F}_p \psi + \mathcal{F}_d \psi,
    \label{eq:neutron_transport}
\end{equation}
where $\psi_g$ is the angular neutron flux in group $g$, $\Sigma_t^g$ is the total macroscopic cross-section. $\mathcal{S}_g\psi$ is the scattering source term,
\begin{equation*}
    \mathcal{S}_g\psi = \sum_{g^\prime=1}^{G}\frac{1}{4\pi}\Sigma_{s}^{g^\prime\to g}\phi_{g^\prime}, \qq{with} \phi_{g^\prime}\qty(\vb{r}) = \int_{4\pi}\dd{\vb*{\Omega}}\psi_{g^\prime}\qty(\vb{r}, \vb*{\Omega})
\end{equation*}
where the scattering has been assumed isotropic. $\phi_{g^\prime}$ is the scalar flux in the energy group $g^\prime$. The prompt and delayed fission term of Eq. \eqref{eq:neutron_transport} are given respectively by:
\begin{equation}
    \mathcal{F}_p \psi = \frac{(1 - \beta)\chi_p^g}{4\pi}\sum_{g^\prime=1}^{G}\nu^{g^\prime}\Sigma_f^{g^\prime}\phi_{g^\prime} \qq{and}\mathcal{F}_p \psi = \frac{(1 - \beta)\chi_p^g}{4\pi}\sum_{g^\prime=1}^{G}\nu^{g^\prime}\Sigma_f^{g^\prime}\phi_{g^\prime},
    \label{eq:fission_source}
\end{equation}
where $\chi^g$ is the fission spectrum, $\nu^{g^\prime}$ is the average number of neutrons produced per fission in group $g^\prime$, $\Sigma_f^{g^\prime}$ is the fission macroscopic cross-section, $\beta = \sum_j \beta_j$ and $\chi_d^g$ is the delayed neutron spectrum. The neutron flux does not directly appear in the delayed neutron source of  Eq. \eqref{eq:fission_source}, but concentrations of DNPs are linear in flux through the source term of Eq. \eqref{eq:precursors_advection_diffusion_RANS_turbulent},
\begin{equation}
    S_j = \frac{\beta_j}{\keff} \sum_{g=1}^{G}\nu^g\Sigma_f^g\phi_g.
    \label{eq:precursors_source}
\end{equation}
In Eqs. \eqref{eq:neutron_transport} and \eqref{eq:precursors_source}, the effective multiplication factor $\keff$ is introduced as $\nu^g \to \nu^g / \keff$ and is playing the role of eigenvalue for the homogeneous problem. A vacuum boundary condition is applied on the angular neutron flux at the top and bottom boundaries of the domain,
\begin{equation}
    \psi^g\qty(\vb{r}, \vb*{\Omega}) = 0, \qq{for} y=0,\ell \qq{and} \vb*{\Omega}\cdot\vu{n} > 0.
    \label{eq:neutron_bc}
\end{equation}
At the inlet and the outlet of the geometry, the angular flux is assumed to be periodic,
\begin{equation}
    \psi^g\qty(\vb{r}, \vb*{\Omega}) = \psi^g\qty(\vb{r} + L\vb*{\Omega}, \vb*{\Omega}), \qq{for} x=0, L
    \label{eq:neutron_bc_periodic}
\end{equation}
A periodic boundary condition is also applied for the DNPs concentration, with a zero flux condition at the top and bottom boundaries of the domain,
\begin{equation}
    \vu{n}\cdot\grad{C_j} = 0, \qq{for} y=0,\ell \qq{and} C_j (0, y) = C_j(L, y).
    \label{eq:precursors_bc}
\end{equation}
\subsection{Angular discretization}\label{subsec:angular_discretization}
The discrete directions on which Eq. \eqref{eq:neutron_transport} is solved are selected in the half-hemisphere due to the symmetry of the problem with respect to the $x-y$ plane. The azimuthal and polar angles, respectively $\varphi$ and $\mu=\cos\theta$ of the direction vector
\begin{equation}
    \vb*{\Omega} = \qty(\sqrt{1-\mu^2}\cos\varphi, \sqrt{1-\mu^2}\sin\varphi),
\end{equation}
are discretized using the Gauss-Legendre and Gauss-Tchebychev quadratures. Any integral of an angular dependent function such as the scalar flux is expressed as,
\begin{equation}
    \phi_g = 2 \int_0^{2\pi}\dd{\varphi}\int_{0}^1 \dd{\mu}  \psi_g\qty(\vb*{\Omega}) = \sum_{n=1}^{N_{\mu}}\sum_{k=1}^{N_{\varphi}} w_n w_k \psi_g\qty(\vb{r}, \vb*{\Omega}_{nk}),
\end{equation}
where $w_n$ and $w_k$ are the weights of the Gauss-Legendre and Tchebychev-Gauss quadratures respectively, and $\vb*{\Omega}_{nk}$ is the direction vector associated to the $n$-th polar angle and the $k$-th azimuthal angle \cite{abramowitz1968handbook}. The number of polar and azimuthal angles are respectively $N_{\theta}$ and $N_{\varphi}$.
\subsection{Space discretization}\label{subsec:space_discretization}
The space discretization of Eq. \eqref{eq:neutron_transport} is performed using a diamond difference scheme \cite{lewis1984computational}. In the following equations, the energy group index is omitted. The angular flux is integrated over the control volume $V_k$,
\begin{equation}
    \psi_{i,j} = \frac{1}{\Delta x \Delta y}\int_{x_{i-1/2}}^{x_{i+1/2}}\int_{y_{j-1/2}}^{y_{j+1/2}}\dd{x}\dd{y}\psi\qty(x,y).
    \label{eq:angular_flux_discretization}
\end{equation}
Advection terms are discretized using the first-order upwind scheme. Along the $x$-axis,
\begin{equation}
    \abs{\Omega_x} \int_{x_{i-1/2}}^{x_{i+1/2}}\int_{y_{j-1/2}}^{y_{j+1/2}}\dd{x}\dd{y}\pdv{\psi}{x} = \abs{\Omega_x}\qty(\psi_{i+1/2,j} - \psi_{i-1/2,j})\Delta y,
\end{equation}
and the same goes for the $y$-axis. Cross-sections are assumed to be constant within the cell. The diamond approximation is used to relate the numerical fluxes at the cell interfaces to the mean angular flux values over a cell \cite{lewis1984computational},
\begin{align}
    \psi_{i+1/2,j} & = 2\psi_{i,j} - \psi_{i-1/2,j}, \qq{if}\vb*{\Omega}\cdot\vu{x} > 0,\label{eq:sweep_left_to_right} \\
    \psi_{i-1/2,j} & = 2\psi_{i,j} - \psi_{i+1/2,j}, \qq{if}\vb*{\Omega}\cdot\vu{x} < 0,\label{eq:sweep_right_to_left} \\
    \psi_{i,j+1/2} & = 2\psi_{i,j} - \psi_{i,j-1/2}, \qq{if}\vb*{\Omega}\cdot\vu{y} > 0,\label{eq:sweep_bottom_to_top} \\
    \psi_{i,j-1/2} & = 2\psi_{i,j} - \psi_{i,j+1/2}, \qq{if}\vb*{\Omega}\cdot\vu{y} < 0.\label{eq:sweep_top_to_bottom}
\end{align}
The boundary condition Eq. \eqref{eq:neutron_bc} is prescribed by setting a zero upwind numerical flux on the boundaries. The periodic boundary condition, Eq. \eqref{eq:neutron_bc_periodic} is enforced by iterating over the $x$-boundaries of the domain during the transport sweep. In Eqs. \eqref{eq:sweep_left_to_right} to \eqref{eq:sweep_top_to_bottom}, $\vu{x}$ and $\vu{y}$ are the unit vectors along the $x$ and $y$ axis, respectively.
\subsection{Iterative scheme}\label{subsec:iterative_scheme}
The steady state neutronic problem defined by Eqs. \eqref{eq:neutron_transport}, \eqref{eq:neutron_bc} and \eqref{eq:precursors_advection_diffusion_RANS_turbulent} is solved using the power iteration method \cite{lewis1984computational}. At the beginning of the calculation, a MOC pre-calculation is performed on the given steady state velocity field to store the \textit{tracking data} needed to solve Eq. \eqref{eq:precursors_advection_diffusion_RANS_turbulent} according to the numerical method defined in Sec. \ref{subsec:calculation_strategy}. The neutron source is decomposed into a self-scattering source, a multigroup scattering source and a fission source,
\begin{equation}
    \qty(\vb*{\Omega}\cdot\grad + \Sigma_t)\psi_g = q_{gg} \qty(\vb*{\Omega}) +  q_{gg^\prime} + \frac{1}{\keff}q_{f}.
\end{equation}
Each contribution is solved iteratively. The self-scattering problem is treated with Richardson fixed-point algorithm, and assumed to be converged when:
\begin{equation}
    \frac{\norm{\phi^{(n+1)} - \phi^{(n)}}}{\norm{\phi^{(n)}}} < \num{1e-5}.
    \label{eq:flux_moments_convergence}
\end{equation}
An additional iterative layer is needed during the transport sweep to account for the periodic boundary condition, Eq. \eqref{eq:neutron_bc_periodic}. For a given direction $\vb*{\Omega}$,
\begin{equation}
    \frac{\norm*{\psi_g^{(n+1)}(0,y,\vb*{\Omega}) - \psi_g^{(n)}(0,y,-\vb*{\Omega})}}{\norm*{\psi_g^{(n)}(0,y,-\vb*{\Omega})}} < \num{1e-5}.
\end{equation}
The multigroup scattering source is treated with the Jacobi method, and is considered converged when the scalar fluxes of all the energy groups reach the convergence criterion of Eq. \eqref{eq:flux_moments_convergence}. The fission source is treated with the power iteration method, and is assumed to be converged when:
\begin{equation}
    \frac{\abs{\keff^{(n+1)} - \keff^{(n)}}}{\keff^{(n)}} < \num{1e-6},
    \label{eq:keff_convergence}
\end{equation}
with an updated multiplication factor given by,
\begin{equation}
    \keff^{(n+1)} = \sqrt{\frac{\int\dd{V} \qty[\qty(1-\beta)\sum_g \nu^g \Sigma_{f}^g \phi_{g}^{(n+1)} + \sum_j \lambda_j C_j^{(n+1)}]^2}{\int\dd{V} \qty[(1-\beta)\sum_g \nu^g \Sigma_{f}^g \phi_{g}^{(n)} + \sum_j \lambda_j C_j^{(n)}]^2}},
\end{equation}
and the $\nu$-fission source calculated at the $(n+1)$-iteration is divided by $\keff^{(n+1)}$. The delayed source is updated at each iteration according to the method presented in Sec. \ref{subsec:calculation_strategy}.
\section{The 2D pipe test case}\label{sec:2D_pipe}
To verify the iterative method developed in Sec. \ref{sec:MMOC}, a 2D pipe at high Reynolds number is considered. The geometry and physical parameters of the fuel salt are given in Sec. \ref{subsubsec:reactor_geometry} while the description of the CFD simulation is described in Sec. \ref{subsec:velocity_diffusivity_fields}. The neutronics parameters are given in Sec. \ref{subsec:neutronics_parameters}. The scaling parameters of Eq. \eqref{eq:precursors_advection_diffusion_RANS_turbulent} are evaluated and commented in Sec. \ref{subsec:scaling_parameters}.
\subsection{System parameters and fuel flow}\label{subsec:geometry_physical_parameters}
\subsubsection{Reactor geometry and fuel salt}\label{subsubsec:reactor_geometry}
The reactor is a rectangle of width $\ell$ and length $L$, with a recirculation loop of size $L$. A fissile fuel salt previously used for MSFR (Molten Salt Fast Reactor) studies is circulating in the reactor \cite{brovchenko2019neutronic,DIRONCO2022111739}. The physical data of the system and fuel salt are prescribed in Table \ref{tab:physical_parameters}.
\begin{table}[hbtp]
    \centering
    \caption{Geometrical parameters of the reactor and thermodynamic parameters of the liquid fuel considered.}
    \vspace{0.5cm}
    \begin{tabular}{ccc}
        \hline
        \hline
        Parameter       & Value                                    & Reference                      \\
        \hline
        $L$             & \SI{2}{\meter}                           & -                              \\
        $\ell$          & \SI{1}{\meter}                           & -                              \\
        $\rho$          & \SI{4.125e3}{\kilo\gram\per\cubic\meter} & \cite{brovchenko2019neutronic} \\
        $\nu$           & \SI{2.46e-6}{\square\meter\per\second}   & \cite{brovchenko2019neutronic} \\
        $\mathrm{Sc}_t$ & 0.5                                      & \cite{DIRONCO2022111739}       \\
        \hline
    \end{tabular}
    \label{tab:physical_parameters}
\end{table}
The turbulent Schmidt number has been chosen to be the most penalizing as possible, i.e. the smallest value available in the literature to increase the diffusivity coefficient \cite{DIRONCO2022111739}. 

The reactor's radial dimensions are intentionally designed to be small, ensuring that precursor diffusion remains significant. For a scaled position \(\vb{r}^\prime \to \alpha \vb{r}\), the advection term \(\vb{u} \cdot \grad{}\) scales as \(\alpha^{-1}\), while the diffusion term \(\div{D_t \grad{}}\) scales as \(\alpha^{-2}\). Consequently, diffusion dominates in smaller systems compared to advection, which is more prominent in larger systems.

This scale effect can also be seen in the dimensionless numbers defined in Eq. \eqref{eq:adim_parameters} as \(\mathcal{A} \propto 1/L\) and \(\mathcal{C}_j \propto 1/L^2\).
\subsubsection{Simulation of an established 2D turbulent flow}\label{subsec:velocity_diffusivity_fields}
\begin{figure}[hbtp]
    \centering
    \includegraphics[width=0.8\textwidth]{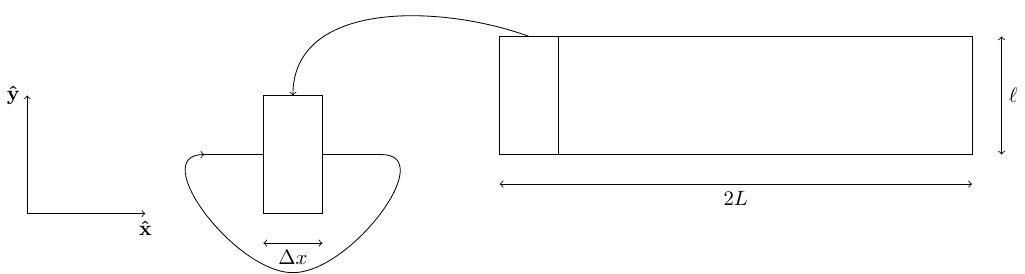}
    \caption{Periodic box used to estimate the inlet flowrate.}
    \label{fig:periodic_box}
\end{figure}

The simulation of the flow was done with TrioCFD using the RANS model \cite{angeli2015overview}. All the balance equations are solved on a Cartesian mesh. As shown in Figure \ref{fig:periodic_box}, the molten salt flows from left to right, with the horizontal sides representing walls. We search to compute a 2D turbulent velocity field for such a configuration without power effect. The fluid is assumed as Newtonian and incompressible thus with a constant density field. Gravity is omitted.  The full geometry is a loop of length \(2L\). The outlet flow distribution is thus applied at the inlet entrance. Thus the flow is fully turbulent developed at the core inlet. As there is no heat and no concentration backward on the flow, it is proposed to evaluate the inlet flow by a periodic box.

The average inlet axial flow velocity is of $u_x = \SI{1}{\meter\per\second}$. The 2D Navier-Stokes equations are solved by the CEA in-house TrioCFD fine CFD code considering the $k-\epsilon$ RANS modeling for turbulence. The boundary conditions are no-slip at walls, constant pressure distribution at outlet, axial fluid velocity, kinetic turbulent energy and turbulent dissipative rate radial distributions at inlet. The Reichardt velocity wall law is assumed at walls and the first fluid velocity point is within the log-part of the universal velocity profile in order to apply correctly the high-Reynolds number $k-\epsilon$ model \cite{hinze1975turbulence}.

In TrioCFD the fully established \(u_x\), \(k\) and \(\epsilon\) radial distributions can be estimated considering a periodic box that mimics, along $\vu{x}$, an infinite \SI{1}{\meter} large rectangle. The periodic box has the same width but a length of \SI{0.01}{\meter} as shown in Fig. \ref{fig:periodic_box}. A mesh by mesh specific force is automatically estimated by TrioCFD through iterations in order to get the inlet flowrate in focus and to respect the pressure inlet/outlet boundary conditions.

\begin{figure}[hbtp]
    \centering
    \includegraphics[width=0.8\textwidth]{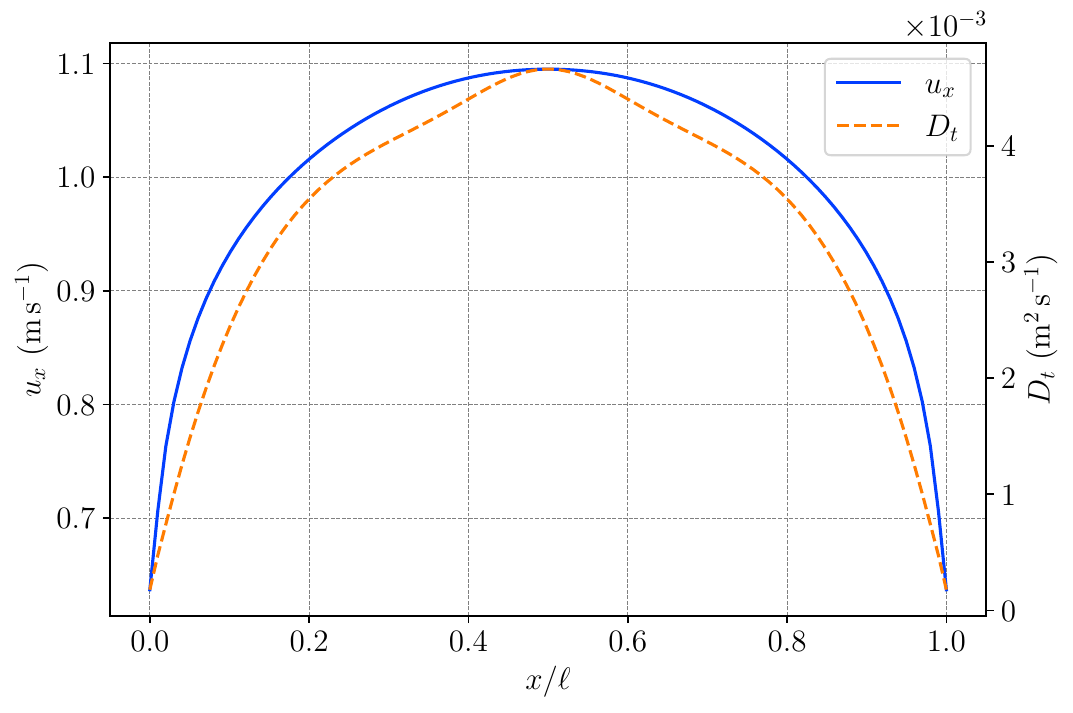}
    \caption{Velocity (\(u_x\)) and turbulent diffusivity (\(D_t\)) radial distributions in the 2D recirculating pipe.}
    \label{fig:velocity_profile}
\end{figure}

Due to the regularity of the 2D rectangle, the inlet radial distributions of \(u_x\), \(k\) and \(\epsilon\) are axially transported. In this case, the pathline tracing presented in Sec. \ref{subsec:pathlines_tracing} is straightforward as all pathlines are parallel to the walls. The radial distribution of turbulent viscosity \(\nu_t\), used in the concentration turbulent diffusion model, is an output of the periodic box computation. Fig \ref{fig:velocity_profile} shows respectively the radial distribution of \(u_x\) and \(D_t\). \(\nu_t\) ranges between \SI{0}{\square\meter\per\second} at walls and \SI{0.001}{\square\meter\per\second}. The maximal value of $\nu_t$ is around 1000 times the fluid kinematic viscosity \(\nu\).
The diffusivity scales with the velocity and is higher at the center of the geometry.
\subsection{Neutronics parameters}\label{subsec:neutronics_parameters}
The nuclear fuel is a fluoride salt previously considered in MSFR studies \cite{laureau2015developpement}. The fissile elements are \ce{^{233}U} and \ce{^{232}Th}. A ten groups energy structure is considered, along with six families of DNPs. To evaluate the boundaries of the energy mesh, a fine energy group calculation in Infinite Homogeneous Medium (IHM) is first performed to estimate the neutron flux per group. The fine flux serves as a weighting factor in the cumulative distribution function (CDF):
\begin{equation}
    \varrho (E, E_0) = \int_{E_0}^{E}\dd{E^\prime} \Sigma_t(E^\prime)\omega(E^\prime)\qq{with} \omega(E) = \frac{\phi(E)}{\int_{0}^{+\infty}\dd{E^\prime}\Sigma_t(E^\prime)\phi(E^\prime)}.
    \label{eq:energy_distribution}
\end{equation}
Eq. \eqref{eq:energy_distribution} calculates the area under the total cross-section with a weight being the flux per units of energy in the IHM. The bounds of the energy group structure are chosen so that the CDF \(\varrho\) has the same value within each group. The fine flux, cross-sections and precursors parameters ($\lambda_j, \beta_j$) are generated for the IHM using OpenMC \texttt{0.14.0} \cite{romano2015openmc}, with \SI{5e3}{} particles per batch and \SI{5e4}{} batches. Precursors values can be found in Table \ref{tab:decay_constants}.
\begin{table}[hbtp]
    \centering
    \caption{Decay constant and delayed neutron fraction of the six DNPs families.}
    \vspace{0.5cm}
    \begin{tabular}{ccc}
        \hline
        \hline
        Family & Decay constant            & Fraction       \\
        \hline
        1      & \SI{1.28e-2}{\per\second} & \SI{2.51e-4}{} \\
        2      & \SI{3.46e-2}{\per\second} & \SI{6.79e-4}{} \\
        3      & \SI{1.19e-1}{\per\second} & \SI{5.45e-4}{} \\
        4      & \SI{2.90e-1}{\per\second} & \SI{1.13e-3}{} \\
        5      & \SI{8.51e-1}{\per\second} & \SI{3.73e-4}{} \\
        6      & \SI{2.60}{\per\second}    & \SI{1.41e-4}{} \\
        \hline
    \end{tabular}
    \label{tab:decay_constants}
\end{table}
\subsection{Scaling parameters}\label{subsec:scaling_parameters}
The scaling parameters of the problem can be calculated for the problem specifications, Sec. \ref{subsec:geometry_physical_parameters} and the DNPs values, Sec. \ref{subsec:neutronics_parameters}. The turbulent diffusivity coefficient is calculated as per Eq. \eqref{eq:turbulent_diffusivity}, $D_t = \SI{4.7e-3}{\square\meter\per\second}$. The Reynolds number of the flow is calculated as \(\text{Re} = u_0 L / \nu = \SI{4.07e5}{}\) and is typical of a turbulent flow. The scaling number $\mathcal{A}$ accounting for the ratio of advection over diffusion is,
\begin{equation}
    \mathcal{A} = \SI{2.15e2}{},
    \label{eq:scaling_number_A}
\end{equation}
which is far lower than what is expected in a laminar flow like the lid driven cavity where only molecular diffusion is taken into account. The scaling number $\mathcal{B}_j$ accounting for the ratio of advection ranges over:
\begin{equation}
    \mathcal{B}_1 = \SI{7.77e1}{} \qq{and} \mathcal{B}_6 = \SI{3.84e-1}{},
    \label{eq:scaling_number_B}
\end{equation}
and spans over two orders of magnitude, showing that long-lived precursors are more affected by advection than short-lived precursors, as expected. The scaling number $\mathcal{C}_j$ accounting for the ratio of turbulent diffusion over advection ranges over:
\begin{equation*}
    \mathcal{C}_1 = \SI{3.62e-1}{} \qq{and} \mathcal{C}_6 = \SI{1.79e-3}{},
    \label{eq:scaling_number_C}
\end{equation*}
showing that turbulent diffusion mostly affects the long-lived precursors.
\section{Results}\label{sec:results}
In this section, the velocity and diffusivity fields used in the coupled neutronic/fluid dynamics simulation are presented in Sec. \ref{subsec:velocity_diffusivity_fields}. These fields are used to compute the DNPs concentration both using the method presented in Sec. \ref{sec:MMOC} and using finite volumes with a given DNPs source. The results of this comparison are presented in Sec. \ref{subsec:verification_finite_volumes}. Once the implementation has been verified, the influence of diffusivity is analyzed in Sec. \ref{subsec:influence_diffusivity} by comparing advection only and advection-diffusion calculations. The differences are presented both on the DNPs concentrations and on the reactivity of the system. In these sections, the fixed point problem is solved using the JFNK method \texttt{scipy.optimize.newton\_krylov} implemented in Scipy 1.14.0. The relative tolerance is set to \num{5e-6} as a lower tolerance does not change the reactivities calculated above the convergence criterion of \num{1e-6} of Eq. \eqref{eq:keff_convergence}.
\subsection{Verification with finite volumes}\label{subsec:verification_finite_volumes}
The implementation of the MOC with diffusivity is verified by comparing the DNPs concentration obtained with the MOC and the finite volume method. The discrepancies between the two methods are presented in Table \ref{tab:verification_finite_volumes}. The error on a calculated quantity $\vb{X}$ between two methods $M_1$ and $M_2$ is defined as \cite{tiberga2020results}:
\begin{equation}
    \epsilon(\vb{X}) = \sqrt{\frac{\sum_{i}\qty(\vb{X}^{M_1}(\vb{r}_i) - \bar{\vb{X}}(\vb{r}_i))^2}{\sum_i \bar{\vb{X}}(\vb{r}_i)^2}}, \qq{with} \bar{\vb{X}}(\vb{r}_i) = \frac{1}{2}\qty(\vb{X}^{M_1}(\vb{r}_i) + \vb{X}^{M_2}(\vb{r}_i)),
    \label{eq:error}
\end{equation}
The error is computed on $\keff$, for each DNPs family and for the total delayed source. Both advection only and advection-diffusion calculations are performed in the comparaison using Eq. \eqref{eq:error}.
\begin{table}[hbtp]
    \centering
    \caption{Comparison of the DNPs concentration obtained with the MOC and the finite volume method.}
    \vspace{0.5cm}
    \begin{tabular}{ccc}
        \hline
        \hline
                               & MOC (adv. only) & MOC (adv. \& diff.) \\
        \hline
        $\keff$                & $\num{1.31e-6}$                   & $\num{4.2e-7}$                            \\
        $C_1$                  & $\num{3.40e-5}$                      & $\num{1.73e-4}$                              \\
        $C_2$                  & $\num{9.14e-5}$                      & $\num{1.95e-4}$                              \\
        $C_3$                  & $\num{3.13e-4}$                      & $\num{3.71e-4}$                              \\
        $C_4$                  & $\num{7.33e-4}$                      & $\num{1.14e-3}$                              \\
        $C_5$                  & $\num{1.63e-3}$                      & $\num{1.65e-3}$                              \\
        $C_6$                  & $\num{1.77e-3}$                      & $\num{1.73e-3}$                              \\
        $\sum_i \lambda_i C_i$ & $\num{1.13e-4}$                      & $\num{2.5e-4}$                              \\
        \hline
    \end{tabular}

    \label{tab:verification_finite_volumes}
\end{table}
In Table \ref{tab:verification_finite_volumes}, the DNPs families with the lowest half-life, i.e. the lower decay constant appear to have the largest error between the two numerical methods. The relative error between the two numerical methods compared, i.e FV (adv. only and adv. plus diffusion) and MOC (adv. only and adv. plus diffusion) is below the percent. The error between the advection only calculations and the advection plus diffusion calculations is of the same order of magnitude. This strengthens the validity of the iterative method presented in Sec. \ref{sec:MMOC}.
\subsection{Influence of diffusivity}\label{subsec:influence_diffusivity}
The influence of diffusivity on the DNPs concentration is analyzed by comparing the results of the advection only and advection-diffusion calculations. The DNPs distributions are presented in Sec. \ref{subsubsec:DNPs_distributions}.
\subsubsection{DNPs distributions}\label{subsubsec:DNPs_distributions}
\begin{figure}[hbtp]
    \centering
    \includegraphics[width=\textwidth]{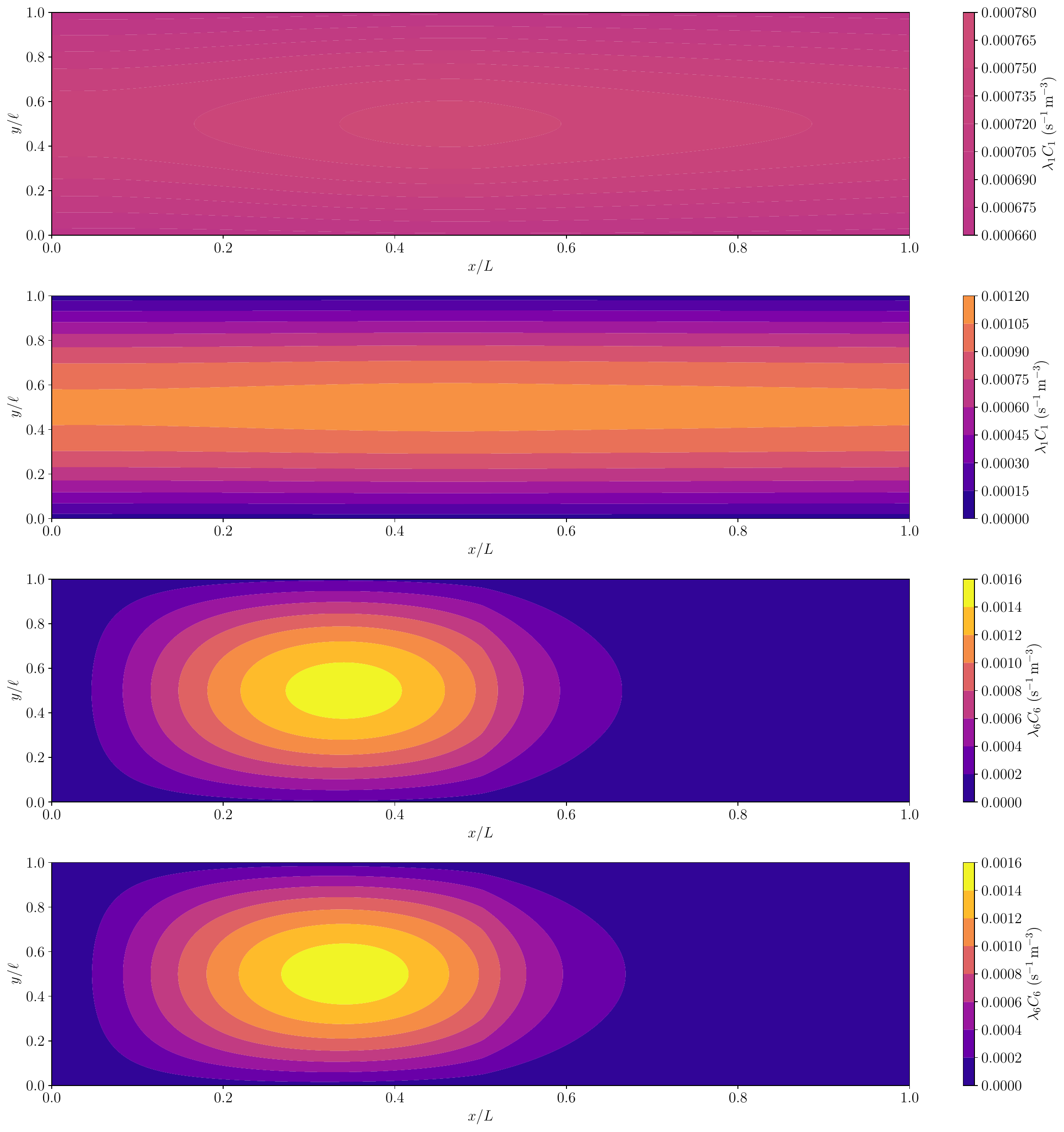}
    \caption{Delayed activity per unit of volume in the reactor, ordered as, from top to bottom, first family with and without diffusivity, sixth family with and without diffusivity, obtained with the MOC method.}
    \label{fig:precursors_concentration_diffusing_MOC}
\end{figure}

\begin{figure}[hbtp]
    \centering
    \includegraphics[width=\textwidth]{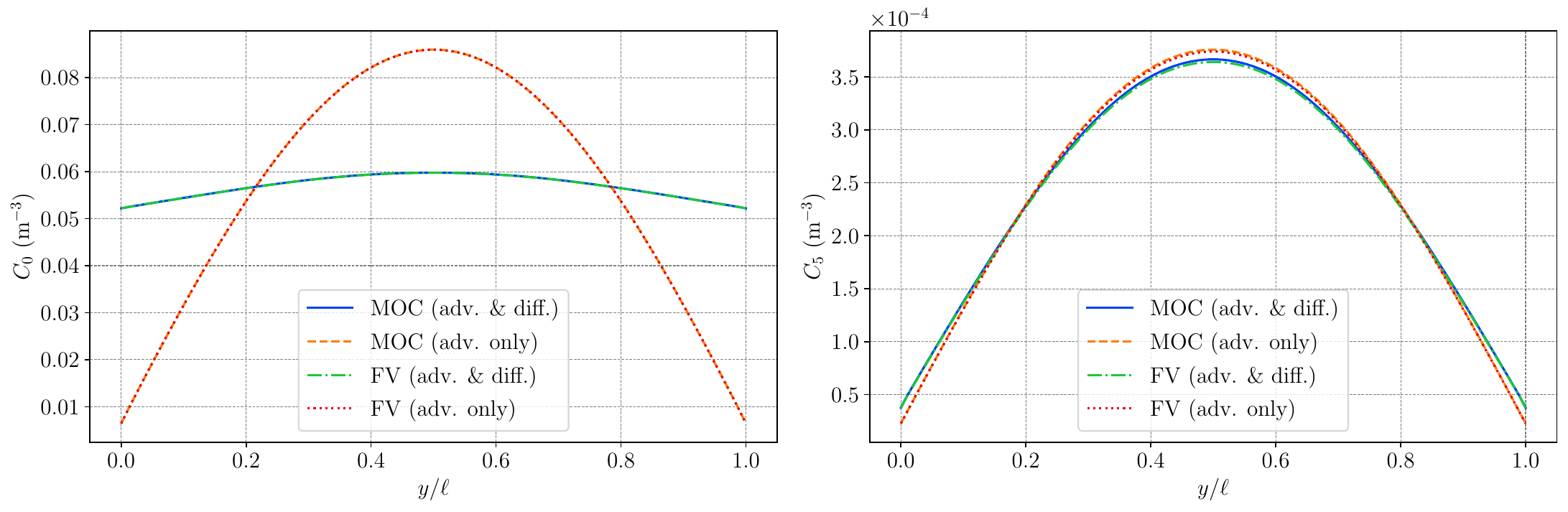}
    \caption{Cross-sections of the DNPs concentration at the outlet of the core, ordered as, first family on the left and sixth family on the right, with and without diffusivity.}
    \label{fig:precursors_concentration_diffusing_FV_cross_sections}
\end{figure}

\begin{figure}[hbtp]
    \centering
    \includegraphics[width=\textwidth]{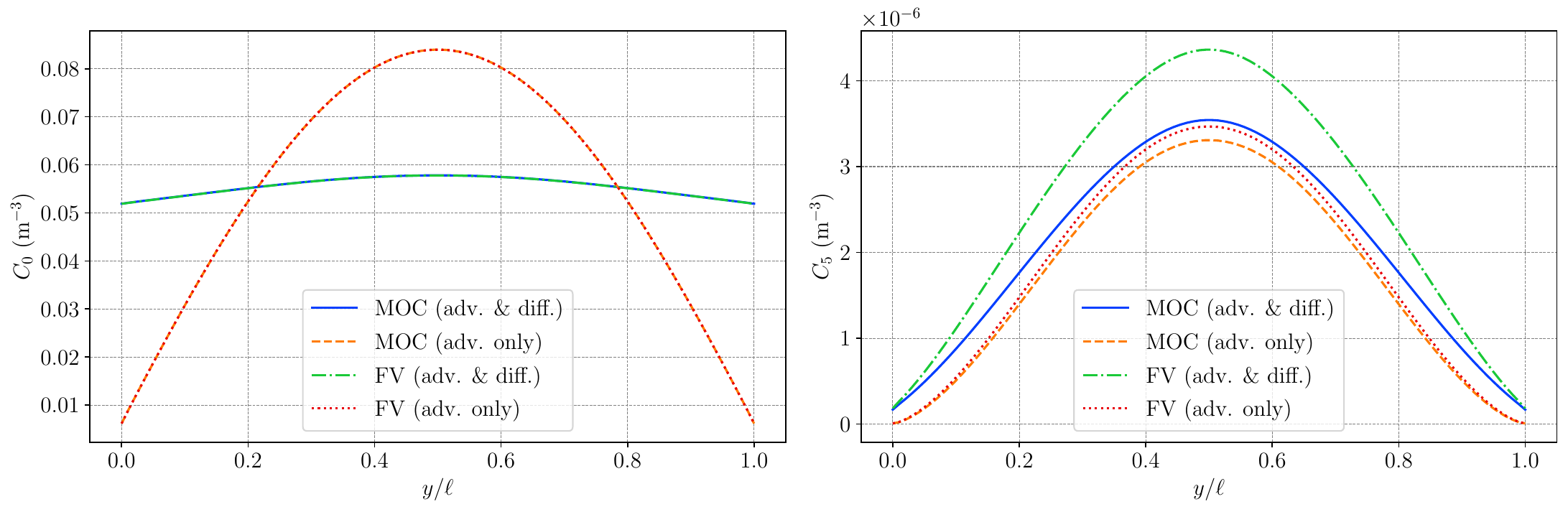}
    \caption{Cross-sections of the DNPs concentration at the outlet of the recirculation loop, ordered as, first family on the left and sixth family on the right, with and without diffusivity.}
    \label{fig:precursors_concentration_diffusing_FV_cross_sections_outlet}
\end{figure}

\paragraph{DNPs distributions in the reactor}
The DNPs distributions of two families, the first and the sixth, Table \ref{tab:decay_constants}, obtained with and without diffusivity using the MOC method of Sec. \ref{subsec:finite_volumes} are presented in Fig. \ref{fig:precursors_concentration_diffusing_MOC}. With advection only, the DNPs concentration of the first family stays almost constant on the straight pathlines crossing the reactor. Considering the sixth family, precursors almost all decay within the core region, and some of them are advected by the stronger flow at the center of the geometry. With diffusivity added to advection, the phenomenon guessed with dimensional analysis and scaling numbers is observed. The long-lived precursors are much more affected by diffusivity and their distribution is almost homogenized with the reactor. The short-lived precursors of the sixth family are less affected even though the peaking concentration seen in advection only is smoothed out. This is an expected result as diffusivity transfers precursors from the center of the core where their concentration is maximal to the edges of the reactor where their concentration is lower due to a lower neutron flux.
\paragraph{Difference between advection only and advection-diffusion}
Concentrations can also be compared between advection only and advection-diffusion calculations. The difference between distributions is calculated using Eq. \eqref{eq:error}. The differences are presented in Table \ref{tab:diffusivity_influence}.
\begin{table}[hbtp]
    \centering
    \caption{Difference between advection only and advection-diffusion calculations.}
    \vspace{0.5cm}
    \begin{tabular}{ccc}
        \hline
        \hline
                               & MOC &  FV \\
        \hline
        $C_1$                  & $\num{3.97e-1}$                           & $\num{3.97e-1}$                         \\
        $C_2$                  & $\num{3.45e-1}$                           & $\num{3.45e-1}$                         \\
        $C_3$                  & $\num{2.26e-1}$                           & $\num{2.26e-1}$                         \\
        $C_4$                  & $\num{1.32e-1}$                           & $\num{1.32e-1}$                         \\
        $C_5$                  & $\num{5.31e-2}$                           & $\num{5.27e-2}$                         \\
        $C_6$                  & $\num{2.05e-2}$                           & $\num{1.93e-2}$                         \\
        $\sum_i \lambda_i C_i$ & $\num{1.87e-1}$                           & $\num{1.98e-1}$                         \\
        \hline
    \end{tabular}
    \label{tab:diffusivity_influence}
\end{table}
The differences between advection only and advection-diffusion calculations are of the order of \SI{1e1}{\percent}. The error is a decreasing function of the decay constant, as expected from the dimensional analysis and scaling numbers, Sec. \ref{subsec:scaling_parameters}. The error on the total delayed source is of the order of \SI{1e1}{\percent}, and is a consequence of the large differences obtained on the long-lived families.
\paragraph{Cross-sections of the DNPs distributions}
The phenomenon seen in Fig. \ref{fig:precursors_concentration_diffusing_MOC} can be further observed by displaying cross-sections of the DNPs distributions along the $y$-axis at the outlet of the core for MOC and FV, Fig. \ref{fig:precursors_concentration_diffusing_FV_cross_sections}.
Long-lived precursors are flattened by diffusivity while short-lived precursors are less affected. The advection-only distributions for both families displayed are similar, indicating that the MOC for DNPs reproduced well the solution obtained using FV. The same distributions can also be displayed at the outlet of the recirculation loop, Fig. \ref{fig:precursors_concentration_diffusing_FV_cross_sections_outlet}.
For long-lived precursors, the distributions are similar to the one obtained at the outlet of the core. For short-lived precursors, some discrepancies appear between both the advection only solutions and the advection-diffusion solutions. The MOC with diffusion tends to under evaluate the concentration of short-lived precursors when this concentration is near zero. The concentration at the outlet of the system is near zero for all methods tested when comparing the short-lived precursors distributions in Fig. \ref{fig:precursors_concentration_diffusing_FV_cross_sections} and Fig. \ref{fig:precursors_concentration_diffusing_FV_cross_sections_outlet}.
\subsubsection{Reactivity Difference}\label{subsubsec:reactivity_difference}
The reactivity differences are calculated for both methods, and for different physical phenomena included in the DNPs balance equation. First, the reactivity difference between the static fuel (no velocity nor diffusivity) and advection-only calculations are,
\begin{equation}
    \Delta \rho (\mathrm{stat/adv})_{\text{FV}} = \ln \qty(\frac{\keff^{\mathrm{adv}}}{\keff^{\mathrm{stat}}}) = \SI{-157.4}{\pcm} \qq{and} \Delta \rho (\mathrm{stat/adv})_{\text{MOC}} = \SI{-157.5}{\pcm}.
    \label{eq:reactivity_difference_static_advection}
\end{equation}
The reactivity loss calculated with the two methods is consistent. The reactivity difference is negative because advection moves the DNPs to a region where the fission cross-section is zero. The reactivity difference obtained in Eq. \eqref{eq:reactivity_difference_static_advection} is roughly equal to $\beta / 2$ because DNPs are produced in a region half the size of the reactor.
The reactivity differences between the advection-only and advection-diffusion calculations are,
\begin{equation}
    \Delta \rho (\mathrm{adv/diff})_{\text{FV}} = \SI{-10.4}{\pcm}, \qq{and} \Delta \rho (\mathrm{adv/diff})_{\text{MOC}} = \SI{-10.3}{\pcm}.
    \label{eq:reactivity_difference_advection_diffusion}
\end{equation}
Again, both the MOC and FV methods give consistent results. The reactivity difference is negative in Eq. \eqref{eq:reactivity_difference_advection_diffusion} because diffusion homogenizes the short-lived DNPs in the reactor, pushing them closer to the boundaries where neutrons are more likely to leak out of the system. The reactivity difference between the advection-only and advection-diffusion calculations is an order of magnitude smaller than the reactivity difference between the static and advection-only calculations. This is expected as advection remains the dominant effect in the system, as seen in Sec. \ref{subsec:scaling_parameters}, and diffusion mostly affects the long-lived precursors.
\subsection{Sensitivity to the turbulent Schmidt number}\label{subsec:sensitivity_turbulent_Schmidt}
As there is no consensus on the value of the turbulent Schmidt number in molten salt \cite{AUFIERO201478, wooten2018, DIRONCO2022111739}, a sensitivity analysis is performed on the reactivity difference between advection and diffusion as a function of the Schmidt number. Varying the turbulent Schmidt number changes the value of $\mathcal{A}$, Eq. \eqref{eq:scaling_number_A}, and therefore alters the strength of diffusive effects relative to advection.
\begin{figure}[hbtp]
    \centering
    \includegraphics[width=0.8\textwidth]{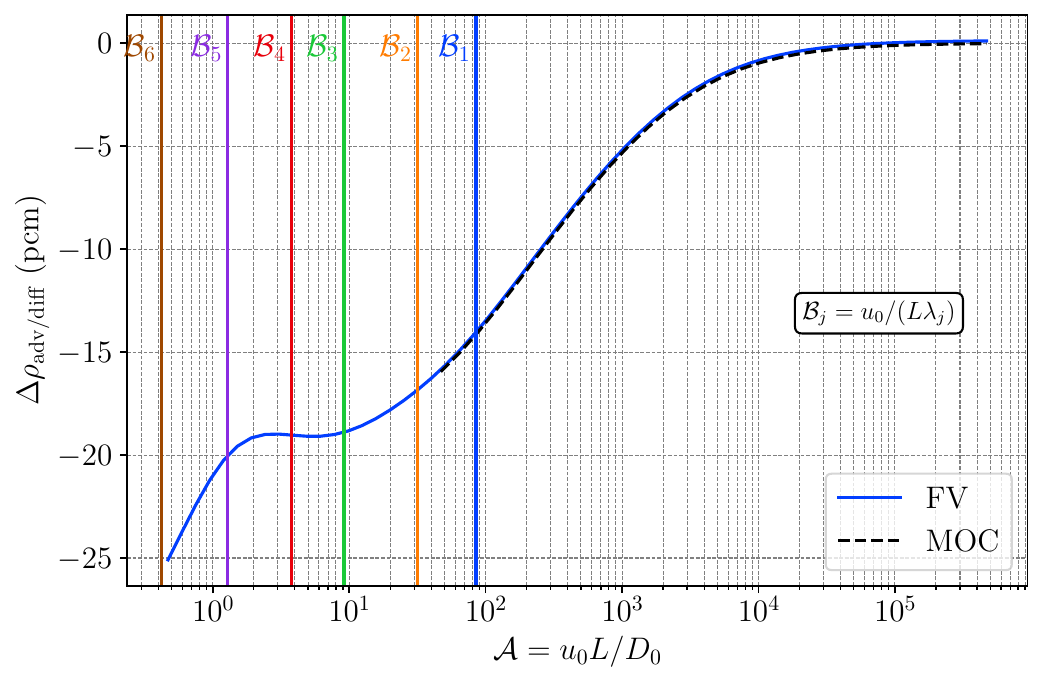}
    \caption{Reactivity difference as a function of \(\mathcal{A}\) by varying the turbulent Schmidt number between \SI{1e-3}{} and \SI{1e3}{}.}
    \label{fig:sensitivity_turbulent_Schmidt}
\end{figure}

In Fig. \ref{fig:sensitivity_turbulent_Schmidt}, the reactivity difference is plotted as a function of the advection-diffusion scaling number \(\mathcal{A}\) (Eq. \eqref{eq:scaling_number_A}) while the vertical lines represent the advection-reaction scaling numbers, \(\mathcal{B}_j\) (Eq. \eqref{eq:scaling_number_B}). For both numerical methods, the reactivity difference decreases as \(\mathcal{A}\) increases, indicating regimes where advection dominates. From Fig. \ref{fig:sensitivity_turbulent_Schmidt}, it is apparent that long-lived precursors are more affected by diffusivity than short-lived precursors, as their scaling number is higher. Consequently, long-lived precursors are the first to become comparable to \(\mathcal{A}\) when diffusion increases. Reaching \(\mathcal{A} \simeq \SI{1e1}{}\), diffusion starts stagnating and then decreases again as new precursors families are getting affected by diffusion.

Both numerical methods gave comparable reactivity differences. However, the iterative MOC method has been stopped at \(\mathcal{A} < \num{1e2}\) as the numerical method took too long to converge.
\section{Discussion and conclusion}\label{sec:conclusion}
The extension of the MOC to include precursor diffusion demonstrated its capability to reproduce the results of the FV method in steady state turbulent flow conditions. The incorporation of diffusivity, which is orders of magnitude higher than molecular diffusivity due to turbulent effects provides a more realistic representation of DNPs behavior. Through scale analysis and simulations, it was shown that turbulent diffusion significantly affects long-lived precursors by homogenizing their distributions more than short-lived precursors. Despite this, the impact on reactor reactivity is minor, as advection continues to dominate the balance equation for delayed neutron precursors (DNPs). Furthermore, the turbulent Schmidt number in molten salts is uncertain, and assuming higher values would further reduce the significance of turbulent diffusion on reactivity. It's also important to note that the system dimensions were chosen to ensure that diffusion remains significant; in larger reactors, the effect of diffusion on reactivity would be even less pronounced.

The new numerical method has been tested on a relatively simple system, specifically a 2D recirculating pipe reactor with high Reynolds number flow and parallel pathlines. Therefore, the performance of the method in more complex systems remains undetermined. Nonetheless, this limitation is also relevant to the standard finite volume (FV) method against which the MOC method is compared. Building and solving the full linear system of equations might not always be feasible. Thus, further testing on a broader range of geometries and flow conditions is crucial to fully assess the robustness and applicability of both methods.

The current implementation employs the \texttt{scipy.optimize.newton\_krylov} fixed point solver. Although effective for the tested system, this solver might become slow or impractical for larger systems with a greater number of unknowns. This scalability issue could limit the method's applicability to real-world, large-scale simulations. More efficient solvers or alternative numerical approaches might be necessary for handling larger problems. Additionally, the numerical method could also suffer from performance issues due to the language in which it is written, potentially requiring further optimization or rewriting in a lower-level language for better efficiency.

The numerical method developed in this work can also be extended to 3D flows and more complex meshes. However, such extensions would require adapting the pathline tracking algorithm and the calculation of diffusive fluxes. This represents a significant challenge and would necessitate further development and testing to ensure accuracy and efficiency.

Finally, the developed method is currently suitable for specific types of turbulence modeling, namely steady-state Reynolds-Averaged Navier-Stokes (RANS) calculations. This restricts its applicability to scenarios where such turbulence models are valid. Extending the method to accommodate other turbulence models would be a significant step towards enhancing its versatility.
\bibliographystyle{unsrt}
\bibliography{bibfile.bib}
\end{document}